\documentclass[twocolumn]{aastex631}

\usepackage{verbatim}
\usepackage{graphicx,url}
\usepackage{amsmath,amssymb}
\usepackage{hyperref}
\hypersetup{colorlinks,
linkcolor=blue, 
urlcolor=magenta, 
anchorcolor=blue,
citecolor=blue
}
\usepackage{subfigure}

\begin{document}

\title{Lensing reconstruction from the cosmic microwave background polarization with machine learning}


\author{Ye-Peng Yan}
\affil{Institute for Frontiers in Astronomy and Astrophysics, Beijing Normal University, Beijing 100875, China; xiajq@bnu.edu.cn}
\affil{Department of Astronomy, Beijing Normal University, Beijing 100875, China}

\author{Guo-Jian Wang}
\affil{School of Chemistry and Physics, University of KwaZulu-Natal, Westville Campus, Private Bag X54001 Durban, 4000, South Africa}
\affil{NAOC-UKZN Computational Astrophysics Centre (NUCAC), University of KwaZulu-Natal, Durban, 4000, South Africa}

\author{Si-Yu Li}
\affil{Key Laboratory of Particle Astrophysics, Institute of High Energy Physics, Chinese Academy of Science, P. O. Box 918-3 Beijing 100049, People’s Republic of China}

\author{Yang-Jie Yan}
\affil{Institute for Frontiers in Astronomy and Astrophysics, Beijing Normal University, Beijing 100875, China; xiajq@bnu.edu.cn}
\affil{Department of Astronomy, Beijing Normal University, Beijing 100875, China}

\author{Jun-Qing Xia}
\affil{Institute for Frontiers in Astronomy and Astrophysics, Beijing Normal University, Beijing 100875, China; xiajq@bnu.edu.cn}
\affil{Department of Astronomy, Beijing Normal University, Beijing 100875, China}

\begin{abstract}
The lensing effect of the cosmic microwave background (CMB) is a powerful tool for our study of the distribution of matter in the universe. Currently, the quadratic estimator (EQ) method, which is widely used to reconstruct lensing potential, has been known to be sub-optimal for the low-noise levels polarization data from next-generation CMB experiments. To improve the performance of the reconstruction, other methods, such as the maximum likelihood estimator and machine learning algorithms are developed. In this work, we  present a deep convolutional neural network model named the Residual Dense Local Feature U-net (RDLFUnet) for reconstructing the CMB lensing convergence field. By simulating lensed CMB data with different noise levels to train and test network models, we find that for noise levels less than $5\mu$K-arcmin, RDLFUnet can recover the input gravitational potential with a higher signal-to-noise ratio than the previous deep learning and the traditional QE methods at almost the entire observation scales.

\end{abstract}

\keywords{
	Cosmic microwave background radiation (322); Observational cosmology (1146); Convolutional neural networks (1938)
}

\section{Introduction}
In observations of the cosmic microwave background (CMB) over the past decades, its anisotropy has been established as a robust and powerful cosmological probe, as shown through many experiments over the past two decades, such as COBE \citep{Mather:1994}, Boomerang \citep{Lange:2001}, WMAP \citep{Bennett:2013}, Planck \citep{Planck:2020}, SPT \citep{Louis:2017}. An important goal for current and next-generation CMB experiments is the precise measurement of anisotropies in its polarization states, particularly the search for the faint primordial polarization B-modes induced by primordial gravitational waves. A lot of upcoming and proposed CMB surveys will be devoted to this direction, such as SPT-3G \citep{Benson:2014}, CMB-S4 \citep{Abazajian:2016}, the Simons Observatory \citep{Ade:2019}, LiteBIRD \citep{Hazumi:2019}, AliCPT \citep{Li:2017}. Gravitational lensing effect of the large-scale structure has a significant impact on CMB observations. Gravitational lensing of the CMB arises from the deflection of CMB photons as they pass through the matter distribution between the surface of the last scattering and us, which will lead to a subtle remapping of the temperature and polarization anisotropies \citep{Lewis:2006}. More specifically, lens smoothes the acoustic peaks of the CMB power spectra and generates non-stationary statistics of CMB fluctuations, and converts E-mode polarization into B-mode polarization \citep{Zaldarriaga:1998,Lewis:2001,Hotinli:2022}. It is therefore essential that we remove the lensing effect (delensing) from the observed CMB to reveal the unaltered primordial signal. On the other hand, CMB lensing is a powerful probe for observing the growth of cosmic structures because it encodes information about the matter fluctuations in the CMB maps observed by our telescopes. Reconstructing and analysing a lensing map from CMB data can help constrain the $\Lambda$CDM cosmological model, the neutrino mass scale,  modified gravity, and a wealth of other cosmological physics \citep{Smith:2006,Planck:2020b}. All in all, reconstruction of the lensing potential and delensing from observable CMB data are key for decoding early universe physics.

CMB photons deflection angle is proportional to the gradient of the lensing potential $\phi$, which is the projected weighted gravitational potential along the line-of-sight between us and the CMB. Quadratic estimator \citep{Zaldarriaga:1999,Hu:2002,Maniyar:2021} has been used with great success to reconstruct the CMB lensing potential from CMB temperature and polarization maps and to detect the effects of gravitational lensing at high significance with existing CMB surveys \citep{Planck:2020b,Wu:2019,Darwish:2021}. 
QE is nearly optimal at current noise level of CMB experiments. However, it will become significantly sub-optimal once the noise level falls below the $\sim 5$ $\mu$K-arcmin \citep{Hirata:2003,Carron:2017,Horowitz:2019}, which could be achieved in the next-generation CMB surveys, such as CMB-S4. Previous works \citep{Hirata:2003,Hirata:2003b,Carron:2017,Carron:2019,Millea:2019,Millea:2020,Legrand:2022} demonstrated that the likelihood-based approaches might significantly enhance the lensing reconstruction by efficiently utilizing higher order statistics of the CMB fields at low noise levels.

Machine learning has demonstrated exceptional capabilities in the field of image processing, such as image recognition and denoising, which has been widely used in astrophysical research \citep{Mehta:2019,Fluke:2020,Petroff:2020,Wang:2022,Yan:2023} as a result of the impressive advancements in computer science in recent years. Machine learning recently demonstrated its potential for CMB lensing reconstruction. Specifically, \cite{Caldeira:2019} and \cite{Heinrich:2022} applied the Residual-Unet (ResUnet) model \citep{Kayalibay:2017,Zhang:2018} to the CMB convergence $\kappa$ reconstruction.  In addition, \cite{Guzman:2021} and \cite{Guzman:2022} used the ResUnet model to reconstruct maps of patchy Reionization and cosmic polarization rotation. In their investigation, this model successfully reconstructs $\kappa$ or rotation maps, and  has lower lensing reconstruction noise than the QE technique at low noise levels. Their results, however, are noise-dependent, and the power spectrum of $\kappa$ field is difficult to reconstruct accurately in the presence of non-negligible noise.  \cite{Li:2022} employs a generative adversarial network (GAN) to precisely reconstruct $\kappa$ power spectrum despite the presence of noise. However, the GAN model's reconstruction of the $\kappa$ map contains more noise than the ResUnet model. As an improvement of the ResUnet model, we introduce a convolutional neural network to reconstruct the CMB lensing convergence field.

This paper is organized as follows. In Section \ref{sec_2}, we briefly review a basic background for the CMB lensing and the quadratic estimator. In Section \ref{sec_3}, We describe the simulated data sets as well as our network model. In Section \ref{sec_4}, we present the results of our method and its comparison with the ResUnet model and quadratic estimator approach. In section \ref{discus} discussions about using the CNN method are presented. Finally, we conclude in Section \ref{sec_5}.

\section{WEAK LENSING OF THE CMB}
\label{sec_2}
In this section, the physical underpinnings of CMB lensing, as well as its construction using the quadratic estimator, will be briefly reviewed. We work in the flat-sky approximation and denote two-dimensional Fourier wave numbers by $\boldsymbol{\ell}$ for CMB fields and $\boldsymbol{L}$ for the lensing potential. Gravitational lensing deflects the path of CMB photons resulting in a remapping of the primary CMB in the sky \citep{Hu:2002,Lewis:2006},
\begin{align}
	&\tilde{T}(\hat{\boldsymbol{n}})=T(\hat{\boldsymbol{n}}+\boldsymbol{\alpha}(\hat{\boldsymbol{n}})), \\
	&(\tilde{Q}(\hat{\boldsymbol{n}}) \pm i \tilde{U}(\hat{\boldsymbol{n}}))(\hat{\boldsymbol{n}})=(Q \pm i U)(\hat{\boldsymbol{n}}+\boldsymbol{\alpha}(\hat{\boldsymbol{n}})),
\end{align} 
where  $\hat{\boldsymbol{n}}$ denotes the line-of-sight direction, ($\tilde{T}$, $\tilde{Q}$, $\tilde{U}$) are the lensed CMB fields, ($T$, $Q$, $U$) are the primordial CMB fields, and $\boldsymbol{\alpha}$ is the deflection angle that describes the remapping. The deflection angle is the gradient of the lensing potential ($\phi$) that is a weighted integral of the gravitational potential ($\psi$) along the line of sight, which can be written as:
\begin{align}
	& \boldsymbol{\alpha} = \nabla \phi, \\
	&\phi = -2\int dD \frac{(\chi_s - \chi)}{\chi \chi_s} \psi(\chi \hat{\boldsymbol{n}},\chi),
\end{align} 
where $\chi$ is the comoving distance along the line-of-sight and $\chi_s$ is the distance to the last-scattering surface. The lensing potential can be also represented by the convergence field $\kappa$ because one can move between the two fields using Poisson equation. In Fourier space, using the flat-sky approximation, this relationship can be written as:
\begin{align}
	\kappa(\boldsymbol{L})& = -\frac{\boldsymbol{L}^2}{2}\phi (\boldsymbol{L}).
\end{align}

Gravitational lensing will affect auto- and cross-spectra of CMB fields and produces correlations between different modes $\ell$, proportional to the lensing potential. In the flat-sky approximation, the off-diagonal mode coupling has the  following form
\begin{align}
	\label{power1}
	\langle \tilde{X}(\boldsymbol{\ell_1})\tilde{Y}(\boldsymbol{\ell_2}) \rangle & \propto f^{\phi}_{XY}(\boldsymbol{\ell_1}, \boldsymbol{\ell_2})\phi(\boldsymbol{L}),
\end{align}
where $\boldsymbol{L}=\boldsymbol{\ell_1}+\boldsymbol{\ell_2}$, ($X$,$Y$) are CMB temperature or polarization fluctuations ($T$,$E$,$B$), and $\langle \rangle$ denotes the average over CMB realization. In this work, we focus on the  quadratic estimator \citep{Hu:2002}, in which the minimum-variance quadratic estimator for the CMB lensing potential is written as 
\begin{align}
	\label{phi2}
	\hat{\phi}_{\beta} &= A_{\beta}(\boldsymbol{\ell}) \int \frac{d^2\boldsymbol{\ell_1}}{(2\pi)^2}X^{\rm obs}(\boldsymbol{\ell_1})Y^{\rm obs}(\boldsymbol{\ell_2})F_{\beta}^{\phi}(\boldsymbol{\ell_1},\boldsymbol{\ell_2}) ,
\end{align}
where the superscript `obs' denotes the observed map including instrumental noise, and $F_{\beta}^{\phi}$ is the  variance filter. $\beta = XY$, where $X,Y \in \{T,E,B\}$. The normalization factor $A^{\phi}_{\beta}(\boldsymbol{\ell})$  is chosen to make the estimator unbiased $\langle \hat{\phi}_{\beta}(\boldsymbol{\ell}) \rangle = \phi(\boldsymbol{\ell})$ and has the following form for the least variance estimator
\begin{align}
	A_{\beta}(\boldsymbol{\ell})& = \left[ \int \frac{d^2\boldsymbol{\ell}_1}{(2\pi)^2} f^{\phi}_{\beta}(\boldsymbol{\ell_1},\boldsymbol{\ell_2})F_{\beta}^{\phi}(\boldsymbol{\ell_1},\boldsymbol{\ell_2})  \right]^{-1},
\end{align}
For the EB estimator, we have
\begin{align}
	\label{phi1}
	F_{EB}^{\phi}(\boldsymbol{\ell_1},\boldsymbol{\ell_2}) & = \frac{f^{\phi}_{EB}(\boldsymbol{\ell_1},\boldsymbol{\ell_2})}{C_{\boldsymbol{\ell_1}}^{EE,\text{obs}} C_{\boldsymbol{\ell_1}}^{BB,\text{obs}}},
\end{align}
and
\begin{align}
	\label{fphi}
	f^{\phi}_{EB}(\boldsymbol{\ell_1},\boldsymbol{\ell_2})  = &\left[ (\boldsymbol{\ell_1} \cdot \boldsymbol{L})C_{\boldsymbol{\ell_1}}^{EE}  - (\boldsymbol{\ell_2} \cdot \boldsymbol{L}) C_{\boldsymbol{\ell_1}}^{BB}\right] \\ 
	&\times \sin2(\varphi(\boldsymbol{\ell_1})-\varphi(\boldsymbol{\ell_2})) \nonumber.
\end{align}
For the EE estimator, we have
\begin{align}
	\label{phi1}
	F_{EE}^{\phi}(\boldsymbol{\ell_1},\boldsymbol{\ell_2}) & = \frac{f^{\phi}_{EE}(\boldsymbol{\ell_1},\boldsymbol{\ell_2})}{2C_{\boldsymbol{\ell_1}}^{EE,\text{obs}} C_{\boldsymbol{\ell_2}}^{EE,\text{obs}}},
\end{align}
and
\begin{align}
	\label{fphi}
	f^{\phi}_{EE}(\boldsymbol{\ell_1},\boldsymbol{\ell_2})  = &\left[ (\boldsymbol{\ell_1} \cdot \boldsymbol{L})C_{\boldsymbol{\ell_1}}^{EE}  + (\boldsymbol{\ell_2} \cdot \boldsymbol{L}) C_{\boldsymbol{\ell_2}}^{EE}\right] \\ 
	&\times \cos2(\varphi(\boldsymbol{\ell_1})-\varphi(\boldsymbol{\ell_2})) \nonumber.
\end{align}
The noise properties of these estimators follow from
\begin{align}
	\langle \hat{\phi}_{\beta}(\boldsymbol{L})\hat{\phi}_{\gamma}(\boldsymbol{L'}) \rangle & = (2\pi)^2\delta(\boldsymbol{L}-\boldsymbol{L'})\left[C_{\boldsymbol{L}}^{\phi \phi}+N^{\phi}_{\beta \gamma}(\boldsymbol{L})\right],
\end{align}
where
\begin{align}
	\label{fphi}
	N^{\phi}_{\beta \gamma}(L)  = & A_{\beta}(L)A_{\gamma}(L) \int \frac{d^2\boldsymbol{\ell}_1}{(2\pi)^2} F^{\phi}_{\beta}(\boldsymbol{\ell_1},\boldsymbol{\ell_2})[F_{\gamma}^{\phi}(\boldsymbol{\ell_1},\boldsymbol{\ell_2})  \\ 
	&\times C_{\ell_1}^{X_{\beta}X_{\gamma}}C_{\ell_2}^{Y_{\beta}Y_{\gamma}}+F_{\gamma}^{\phi}(\boldsymbol{\ell_1},\boldsymbol{\ell_2})C_{\ell_1}^{X_{\beta}Y_{\gamma}}C_{\ell_2}^{Y_{\beta}X_{\gamma}}]   \nonumber.
\end{align}

\begin{figure*}
	\begin{center}
		\includegraphics[width=0.9\hsize]{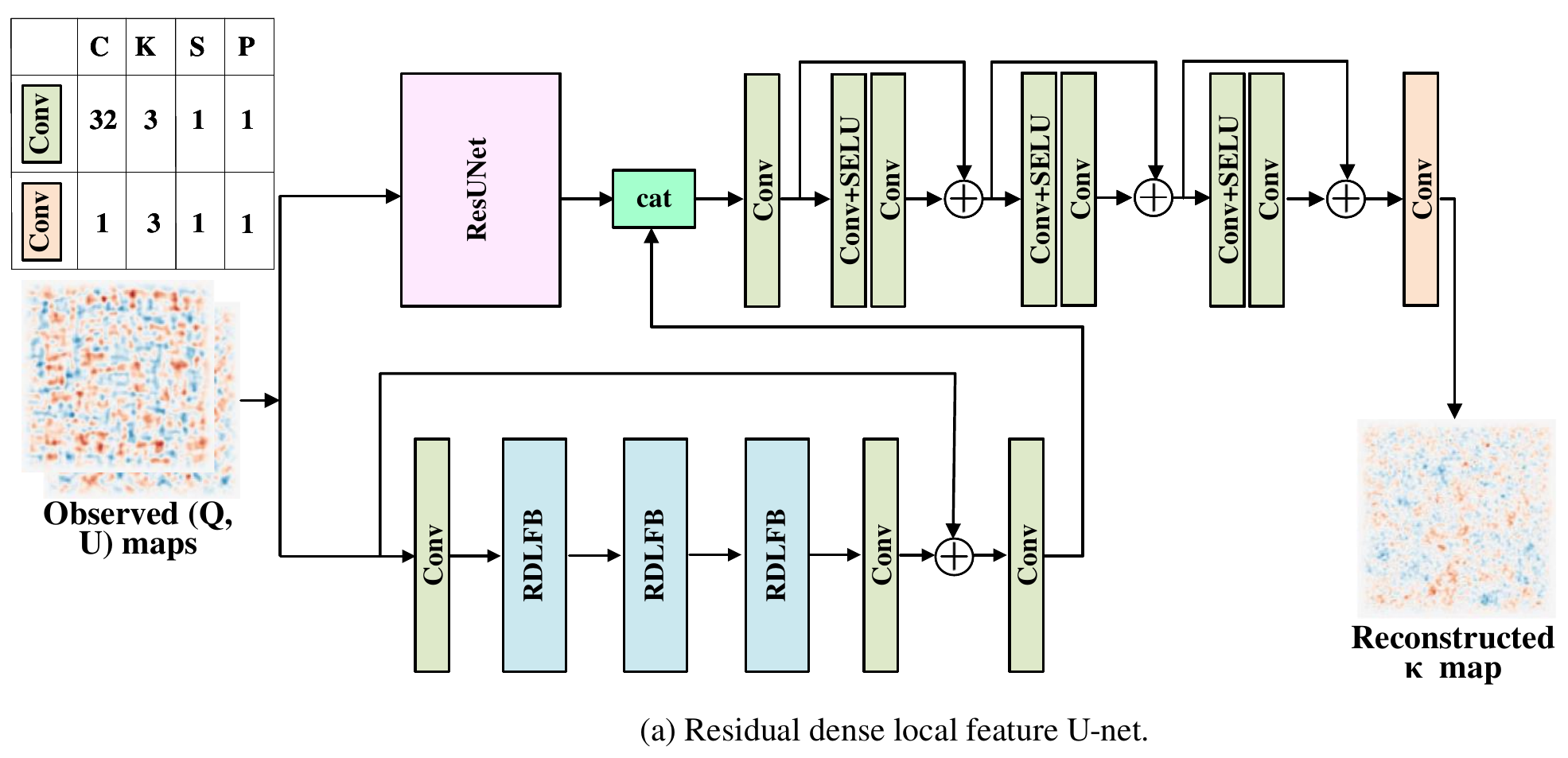}
		\includegraphics[width=0.9\hsize]{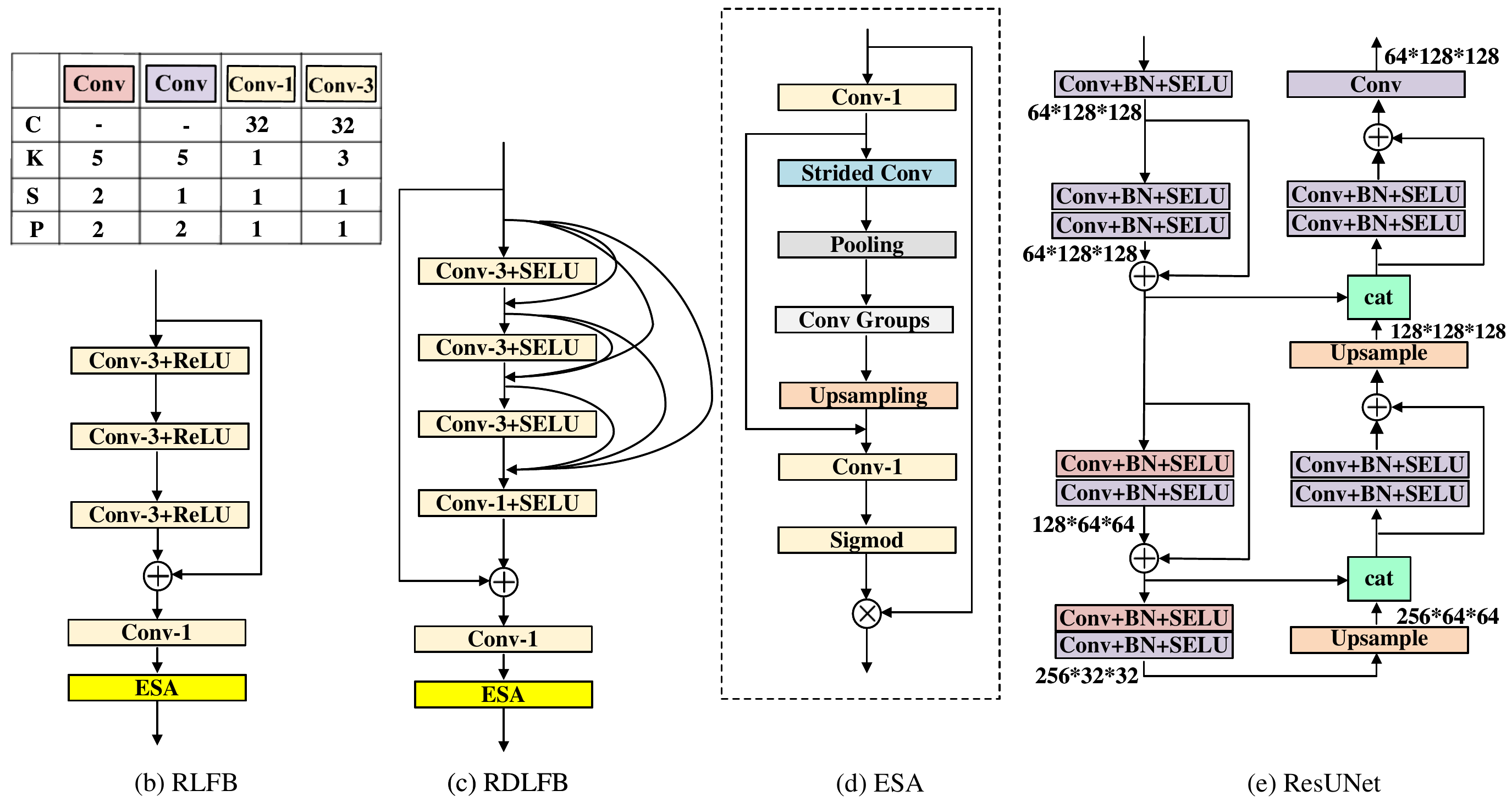}
	\end{center}
	\vspace{-0.4cm}
	\caption{(a) RDLFUnet: The architecture of residual dense local feature U-net. Conv and $\bigoplus$ represent the convolution layer and summation. The cat denotes concatenation operation along the channel dimension. C, K, S, and P represent the number of output channels, kernel (filter) size,  stride, and padding of the convolutional layer, respectively. Bottom panels: BN, SELU, and $\bigotimes$ represent  batch normalization, Scaled Exponential Linear Unit (SELU) function, and multiplication, respectively.  (b) RLFB: residual local feature block from \citet{Kong2022}. (c) RDLFB: the residual dense local feature block. (d) ESA: Enhanced Spatial Attention from \citet{Kong2022}. (e) ResUnet. The shapes (channel, height, width) of feature maps are shown in the panel (e).}
	\label{figure_net}
\end{figure*}

Similarly to \cite{Heinrich:2022}, we use in this paper the minimum variance combination of all the polarization pairs,
\begin{align}
	\hat{\phi}^{QE}(\boldsymbol{L})& = \sum_{\beta}\omega_{\beta}(L)\hat{\phi}^{QE}(\boldsymbol{L}),
\end{align}
instead of the EB-only estimator used in \cite{Caldeira:2019}. Here, the minimum variance weighting is 
\begin{align}
	\omega_{\beta}& = \frac{\sum_{\gamma}(N^{-1})_{\beta \gamma}}{\sum_{\gamma \eta}(N^{-1})_{\gamma \eta}}.
\end{align}
The reconstruction noise of the minimum-variance estimator is then
\begin{align}
	\label{qe_NL}
	N_L& = \frac{1}{\sum_{\beta \gamma}(N^{-1})_{\beta \gamma}}.
\end{align}



\section{METHODOLOGY}
\label{sec_3}
In this section, we will introduce the deep learning architecture we use in this work and the details of the data pipeline and network training.
\subsection{Network architecture}
The convolutional neural network (CNN) has been widely applied to image processing tasks. Its power relies on the network architecture that consists of a stack of non-linear parametric models. By training, the network architecture can convert complex problems into parameter optimization. The convolutional layer \citep{Dumoulin:2016} is the core of the network architecture. Each convolutional layer accepts the feature images from the former layer as an input, then convolves them with a bank of local spatial filters (or kernels) whose values are parameters to be learned, and finally gives output to the next layer after a nonlinear activation function. The output size of a convolutional layer is controlled by three hyper-parameters: the number of output channels (the number of convolutional filters), stride, and amount of zero padding. The stride is defined as the distance in pixels between the centers of adjacent filters. Residual connection \citep{He:2015}, which widely used in image processing tasks, is designed to learn the residual mapping by adding inputs of a given layer to the outputs of the layer that one. The U-Net  \citep{Ronneberger:2015} is a successful architecture that takes an encoder–decoder with convolutional layers and adds extra shortcuts (skip connections) between the encoding and decoding layers to allow for propagation of small-scale information that might be lost when the size of the images decreases. In this work, we provide Residual Dense Local Feature U-Net (RDLFUnet), a deep learning architecture designed to reconstruct the CMB lensing potential. This network architecture combines Residual-Unet (ResUnet) \citep{Kayalibay:2017,Zhang:2018} and the modified Residual Local Feature Network (RLFN) \citep{Kong2022}. The ResUnet is created in U-Net using the residual connections. The RLFN is an excellent architecture for image super resolution reconstruction. 

\begin{figure*}
	\begin{center}
		\includegraphics[width=0.9\hsize]{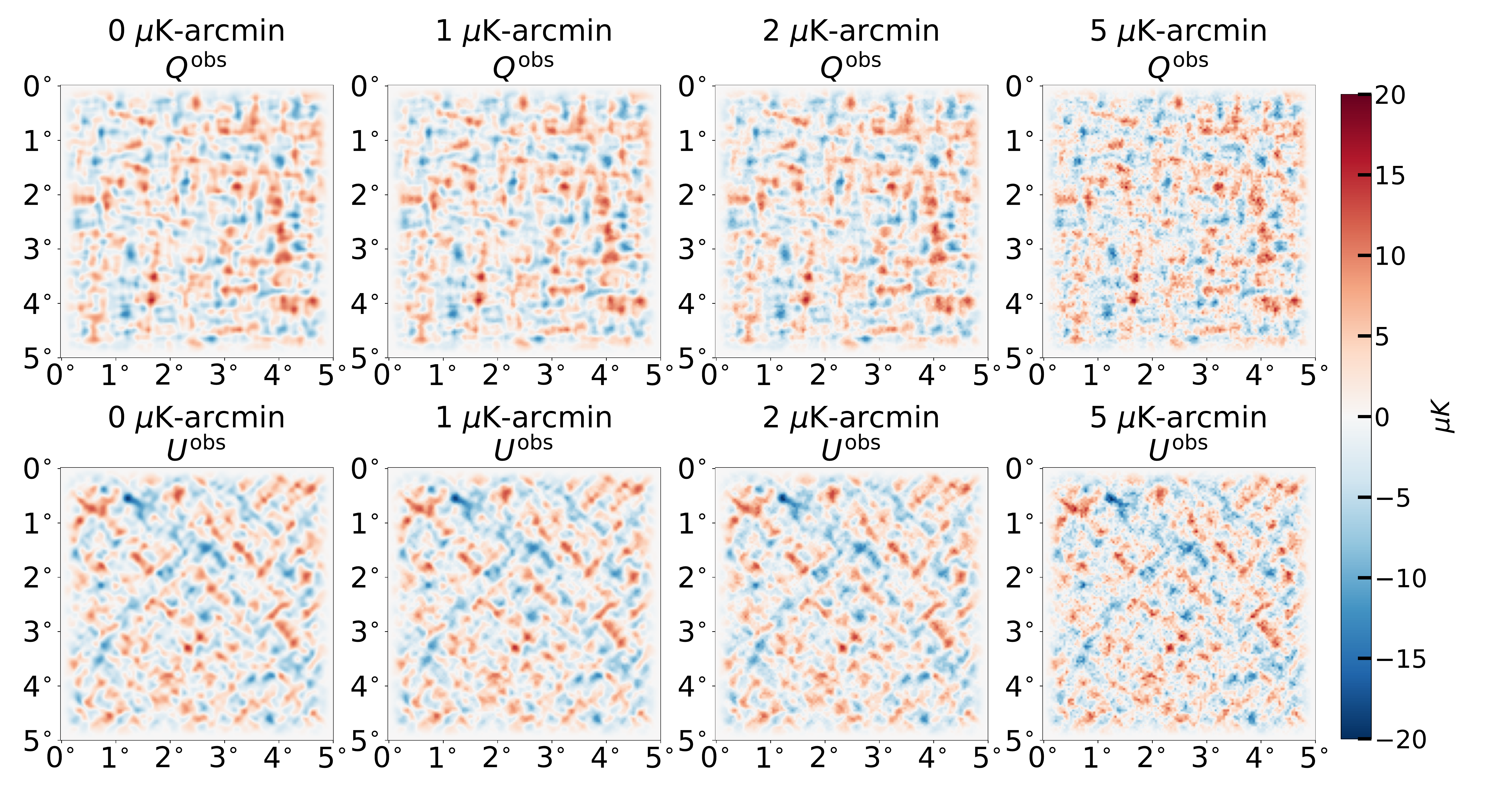}
	\end{center}
	\vspace{-0.4cm}
	\caption{Example of the observed Q (top) and U (bottom) maps with different noise levels.}
	\label{input_map}
\end{figure*}

Our network architecture is divided into two branches as shown in Figure \ref{figure_net}. The first branch is a modified RLFN. The residual local feature block (RLFB), which employs three convolutional layers for residual local feature learning, is the fundamental building block of the RLFN as depicted in the bottom panel (b). The RLFB finally employs an Enhanced Spatial Attention (ESA) block \citep{Liu:2022} to produce the end output. The ESA block can make the residual features to be more focused on critical spatial contents. In this work, the residual dense local feature block (RDLFB), which is depicted in the bottom panel (c), replaces the RLFB. In contrast to the RLFN with residual connection, the RDLFB uses a residual dense connection \citep{Huang:2016,Zhang:2018b}. In our test, such a change might enhance the ability to reconstruct CMB lensing potential. The second branch, known as ResUnet, is shown in the bottom panel (e) of Figure \ref{figure_net}. ResUnet encodes relevant information from the input maps into smaller maps, and then decodes that information to generate the output maps. The output of these two branches is concatenated along the channel dimension, then pass to three residual blocks, and the gravitational convergence $\kappa$ map will be finally constructed. Our network model is built in \texttt{PyTorch}\footnote{\url{https://pytorch.org/}} environment, which is an open-source optimized tensor library for deep learning.

\begin{figure*}
	\begin{center}
		\includegraphics[width=1\hsize]{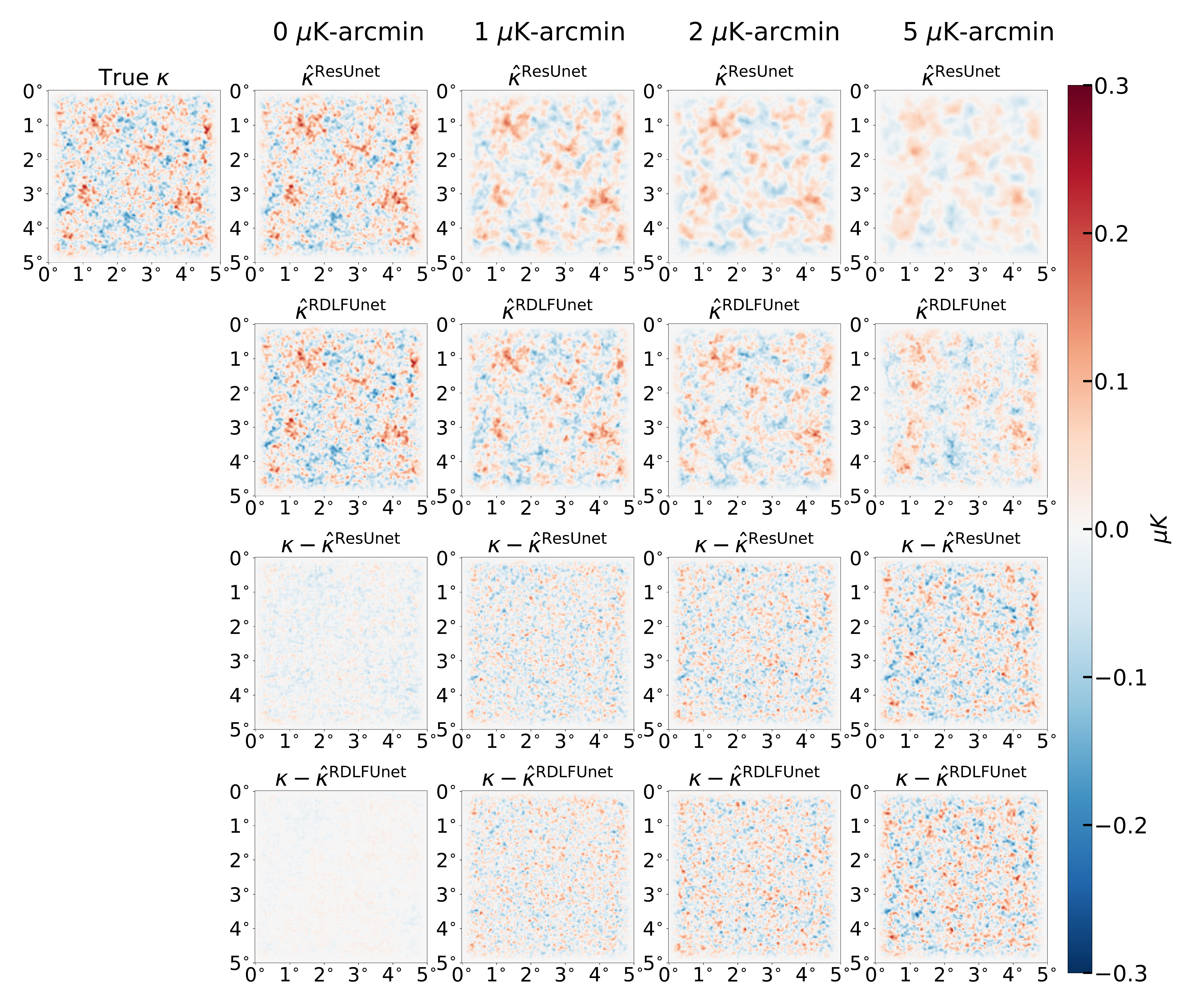}
	\end{center}
	\vspace{-0.4cm}
	\caption{Example of the reconstructed $\hat{\kappa}$ for four noise levels (0, 1, 2, 5 $\mu$K-arcmin; left to right). The top-left map is the ground truth $\kappa$. The ResUnet \citep{Caldeira:2019} and RDLFUnet predictions of $\hat{\kappa}$ are shown in the top and second rows. The related residual maps $\kappa - \hat{\kappa}$  are shown in the third and fourth rows.}
	\label{results_map}
\end{figure*}
The network model is optimized by minimizing a loss function.  The training set consist of $S$ pairs of samples $\{x_i, y_i\}_{i=1}^S$, where $x$ represents  the observed Q/U maps, and $y$ is the corresponding ground truth of $\kappa$ map. Our loss function defines as 
\begin{align}
	\label{loss}
	\mathcal{L} = \mathcal{L}_{\rm LAD} + \beta \mathcal{L}_{\rm FFT},
\end{align}
where $\mathcal{L}_{\rm LAD}$ is the least absolute deviation (LAD, also called L1 loss), $\mathcal{L}_{\rm FFT}$ is the Fourier space loss, and $\beta$ is a coefficient representing the contribution of $\mathcal{L}_{\rm FFT}$ to the total loss and we set $\beta=1$ throughout the paper. The L1 loss has form
\begin{align}
	\mathcal{L}_{\rm LAD}&=\frac{1}{N}\sum_{n=1}^{N}\left[ \frac{1}{WH}\sum_{w=1}^{W}\sum_{h=1}^{H}(|I^{n}_{w,h}-y^{n}_{w,h}|)\right],
\end{align} 
where $N$ is the batch size, $H$ ($W$) is the height (width) of the images in pixels, and $I = f(x)$ ($f(\cdot)$ is the network model) is the predicted image. The form of $\mathcal{L}_{\rm FFT}$ is defined as 
\begin{align}
	\mathcal{L}_{\rm FFT}&=\frac{1}{N}\sum_{n=1}^{N}\left[ \frac{1}{WH}\sum_{w=1}^{W}\sum_{h=1}^{H}(|A_{\rm F}(I^{n}_{w,h})-A_{\rm F}(y^{n}_{w,h})|)\right],
\end{align}
where $A_{\rm F}$ is the amplitude of FFT, which has the form of 
\begin{align}
	A_{\rm F}(I)& = \sqrt{Re[{\rm FFT}(I)]^2+Im[{\rm FFT}(I)]^2},
\end{align}
where $Re[\cdot]$ and $Im[\cdot]$ denote the real and imaginary part, respectively.

\subsection{Data pipeline and network training}
\label{data_pipline}
As a supervised machine learning technique, the CNN method needs a training dataset with values that are already known to be true. Our data set is based on a simulation of a flat sky. First, the publicly available \texttt{CAMB}\footnote{\url{https://github.com/cmbant/CAMB}} package is used to calculate the primordial CMB and lensing power spectra \citep{Lewis:2000}. Here, we use a standard $\Lambda$ cold dark matter model with parameters ($H_0, \Omega_bh^2, \Omega_ch^2, \tau, A_s, n_s$), and their best-fit value and standard deviation can be obtained from the Planck 2015 data \citep{Planck:2016}. Then, using the theoretical power spectra from \texttt{CAMB}, we simulate two-dimensional maps that cover a $5^{\circ} \times 5^{\circ}$ patch of sky with $128 \times 128$ pixels by using a modified version of \texttt{Orphics}\footnote{\url{https://github.com/msyriac/orphics}} and \texttt{resunet-cmb}\footnote{\url{https://github.com/EEmGuzman/resunet-cmb}}.

Previous works \citep{Caldeira:2019,Guzman:2021} employed a fixed lensing power spectrum to generate observed CMB map, however they indicated that the trained network model is sensitive to the variations in cosmological parameters. \cite{Li:2022} used a varying lensing power spectrum  $\alpha C_{\ell}^{\kappa \kappa}$ ($\alpha$ is random values within range $0.75-1.25$), while the cosmological parameters remained constant. In this work, in order to be more realistic and eliminate the dependence of the cosmological parameters, we treat the values of cosmological parameters as independent Gaussian random variables, where the mean values and standard deviations are taken from the best-fit values and standard deviation of Planck-2015 results \citep{Planck:2016}.

Finally, similar to \cite{Caldeira:2019}, \cite{Guzman:2021} and \cite{Li:2022}, we smooth the maps with a Gaussian beam size of FWHM= 1 acrmin and choose 4 levels of noise  with $0.0$ $\mu$K-arcmin,  $1.0$ $\mu$K-arcmin, $2.0$ $\mu$K-arcmin, $5.0$ $\mu$K-arcmin. Each generated map applies a cosine taper of $0.5^\circ$ to reduce edge effects. In total five different types of maps are generated, ($Q^{\rm prim}$,\,$U^{\rm prim}$,\,$Q^{\rm obs}$,\,$U^{\rm obs}$,\,$\kappa$). Here, the superscript `prim' represents the primordial CMB map, while `obs' denotes the observed CMB map including the lensing and instrumental information. 
To generate the observed maps ($Q^{\rm obs}$,\,$U^{\rm obs}$), the primordial maps ($Q^{\rm prim}$,\,$U^{\rm prim}$) are first lensed with the convergence map, then a noise map is added after these lensed maps have been smoothed with a Gaussian beam. The example observed CMB maps with varying noise levels are shown in Figure \ref{input_map}. The variations between noise levels of 0 and $>1 \mu$K-arcmin are readily visible on the maps.

For each  noise level, 25000 sets of the five maps described above are generated and  split into training, validation, and test sets with a ratio of 8:1:1.  Each noise level is trained using a separate network. To train the network, mock observed CMB maps ($Q^{\rm obs}$,\,$U^{\rm obs}$) will be fed into the network model, going through each level, and outputting a $\hat{\kappa}$ map. We use a batch size of 32, adopt the Adam optimizer \citep{Kingma:2014}, and initially set the learning rate to 0.005, which gradually drops to $10^{-6}$ during the iteration. Iterations totaling 60,000 are used to train the network model. We used two NVIDIA Quadro GV100 GPUs to train network, and one network model requires $\sim $12 hr to train.

\section{Results}
\label{sec_4}
Results for the reconstructed $\kappa$ map using our network and its power spectra are presented in this section. We will compare our results to those of the ResUnet model \citep{Caldeira:2019,Guzman:2021}. The reconstructed map is represented by $\hat{\kappa}$, whereas the true map is represented by $\kappa$.

\subsection{Reconstructed map} 
First, we apply our network to the task of $\kappa$ map reconstruction. We take  the observed $Q^{\rm obs}$ and $U^{\rm obs}$ maps as input, and the desired output is the ${\kappa}$ map. The test set is fed into the network after the network training is finished. The reconstructed $\hat{\kappa}$ maps are shown in Figure \ref{results_map} for the four noise experiments. In the case of noiseless input, the residual map contains minimal information, implying that the network can recover the $\kappa$ map reasonably effectively. However, as noise levels increase, all residual map amplitudes increase, and some small-scale structure details are missed by the reconstructed $\kappa$ maps.  Despite the presence of noise, our network can also recover the most large-scale structure of the $\kappa$ map.

To compare with the ResUnet model, we train the ResUnet model separately for each noise level. For the ResUnet training, we use the same training pipeline and network architecture as \citet{Caldeira:2019}. In Figure \ref{results_map}, we compare the results of the ResUnet with our network model at the level of the reconstructed map. We can see that, for noiseless input, the residual map  $\kappa-\hat{\kappa}$ has a lower amplitude than ResUnet's. We also calculate the structural similarity index measure (SSIM) \citep{Zhou:2004} of the reconstructed and true $\kappa$ map. The SSIM is a method used for measuring the similarity between two images, as follows 
\begin{align}
	\label{SSIM}
	{\rm SSIM}(x,y) = \frac{(2\mu_x \mu_y +c_1)(2\sigma_{xy}+c_2)}{(\mu_x^2+\mu_y^2+c_1)(\sigma_x^2+\sigma_y^2+c_2)},
\end{align}
here, $x$  and $y$ represent two images, $\mu_x$ and $\mu_y$ are the pixel sample mean of $x$ and $y$, $\sigma_x$ and $\sigma_y$ are  the variance of $x$ and $y$, $\sigma_{xy}$ is the covariance of $x$ and $y$, and $c_1$ and $c_2$ are two  constants  to stabilize the division with weak denominator. The two images are more similar the closer the value of SSIM is to 1. The two images are precisely the same if its value is 1. For the noiseless input case, the SSIMs are $0.98\pm 0.003$ for our model and $0.89\pm 0.015$ for the ResUnet model, indicating that our reconstructed $\hat{\kappa}$ map is closer to the true $\kappa$ than result from ResUnet model. 

When considering the non-zero noise inputs, from visual inspection, our reconstructed $\hat{\kappa}$ maps can also have more small-scale structure features than the ResUnet model. Compared to ResUnet model, our network can capture more details on small-scale structure. We also calculate the SSIM of the reconstructed and true $\hat{\kappa}$ map. The SSIMs of our network  for the  $1,\,2,\,5\,\mu$K-arcmin noise levels experiments are $0.68\pm 0.010$, $0.56\pm 0.012$ and $0.41\pm 0.015$, while the ResUnet model are $0.62\pm 0.015$, $0.52\pm 0.014$ and $0.39\pm 0.017$, respectively. For experiments of three noise levels, our model performs better than the ResUnet model. 


\begin{figure}
	\begin{center}
		\includegraphics[width=1\hsize]{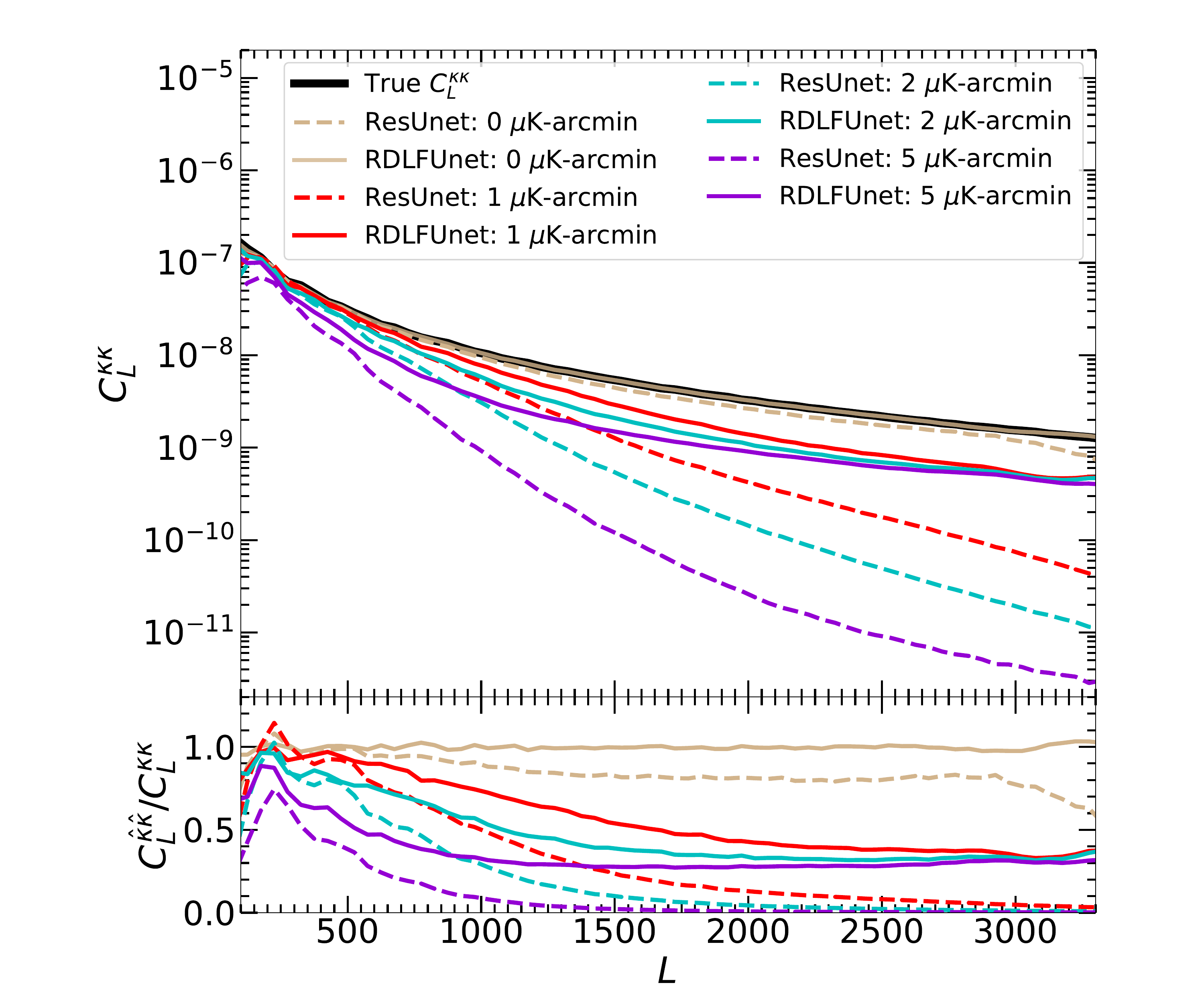}
	\end{center}
	\vspace{-0.4cm}
	\caption{Power spectra of predicted $\hat{\kappa}$ map from the ResUnet (dashed line) and RDLFUnet (solid line) for four noise levels (0, 1, 2, 5 $\mu$K-arcmin) averaged over all realizations in the test set, and the length of each $L$ bin is set to be 50 here. The ratio of the predicted power spectrum to the true power spectrum is also shown.} 
	\label{results_po}
\end{figure}

\subsection{Reconstructed power spectrum} 
Next, we present the power spectrum of the reconstructed $\hat{\kappa}$ map at each noise level and compare them to the results of the ResUnet and QE approaches. 

Figure \ref{results_po} shows the power spectra of reconstructed $\hat{\kappa}$ for four input noise levels. The power spectra are the mean result over the testing set. For the noiseless input, the constructed $\kappa$ power spectra from our network is quite consistent with the target truth power spectrum. However, as noise levels rise, the reconstructed power spectrum visibly degrades at small scales, which  implies the small-scale fluctuations are suppressed. The ResUnet model's reconstructed power spectra for each noise level are also shown in figure \ref{results_po} for comparison's sake. At all noise levels, we can clearly observe that $C_{L}^{\hat{\kappa}\hat{\kappa}}$ reconstructed by our model contains more small-scale information than the ResUnet model, which is consistent with the reported results of the reconstructed $\hat{\kappa}$ map in Figure \ref{results_map}.

\begin{figure}
	\begin{center}
		\includegraphics[width=1\hsize]{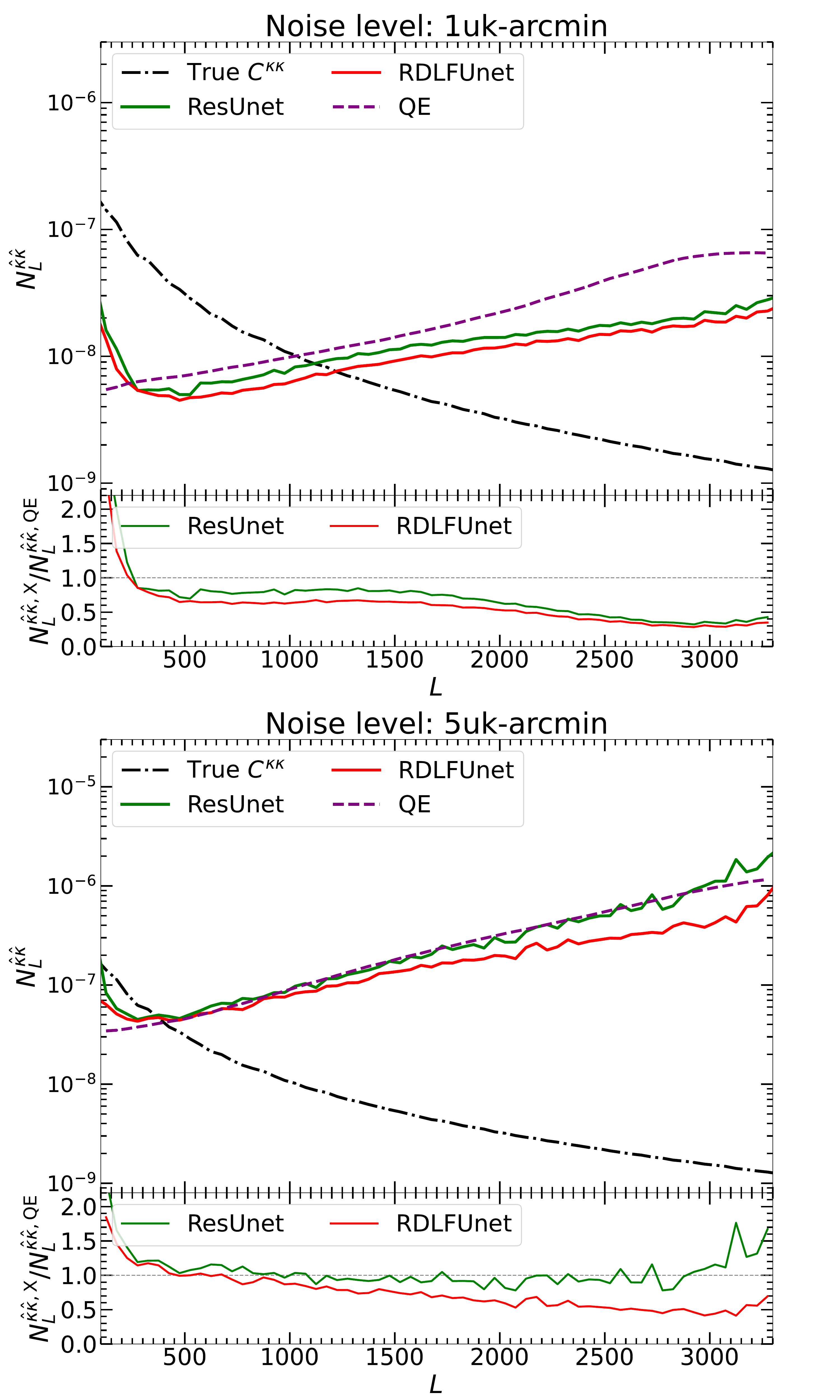}
	\end{center}
	\vspace{-0.4cm}
	\caption{Reconstruction noise power spectra of the ResUnet model, the quadratic estimator, and our model RDLFUnet for noise levels of $1\ \mu$K-arcmin (top panel) and $5\ \mu$K-arcmin (bottom panel). `X' denotes ResUnet or RDLFUnet.}
	\label{recons_noise_power_}
\end{figure}

Finally, we compare the performance of the quadratic estimator in reconstructing $\hat{\kappa}$ power spectrum to that of our network model.  First we need to define the reconstructed noise. As shown in Figure \ref{results_po} and Refs. \citep{Caldeira:2019,Guzman:2021}, the reconstructed $\kappa$ power spectrum $C^{\hat{\kappa}\hat{\kappa}}$ by network model is a bias result. We use a method from \citet{Caldeira:2019} to treat this bias. We normalize the biased $\hat{\kappa}$ field by a factor $R_{L}$ defined as 
\begin{align}\label{Rnorm}
	R_{L}&=\left[ \frac{\langle C_{L}^{\kappa\hat{\kappa}}\rangle}{\langle C_{L}^{\kappa \kappa} \rangle} \right]^{-1},
\end{align} 
where $\langle...\rangle$ represents the average over the test set. It should be noted that $R_{L}$ is a function of scales $L$. Finally, the equivalent of the  unbiased $\kappa$ field is given by 
\begin{align}
	\label{norm}
	\hat{\hat{\kappa}}_L&=R_{L} \hat{\kappa}_{L}.
\end{align}
We assume that $\hat{\hat{\kappa}}$ can be modelled as $\hat{\hat{\kappa}}=\kappa + n$, where $n$ is an uncorrelated noise term. The equivalent of the reconstruction noise spectrum can be defined as 
\begin{align}
	N^{\kappa \kappa}_L&=\langle C^{\hat{\hat{\kappa}} \hat{\hat{\kappa}}}_{L} \rangle - \langle C^{\kappa \kappa}_{L} \rangle,
\end{align}
where $C^{\kappa \kappa}_{L}$ is true power spectrum, and we have
\begin{align}
	\label{rescale}
	\langle C^{\hat{\hat{\kappa}} \hat{\hat{\kappa}}}_{L} \rangle&=R_{L}^2\langle C^{\hat{\kappa} \hat{\kappa}}_{L} \rangle.
\end{align}

\begin{figure}
	\begin{center}
		\includegraphics[width=1\hsize]{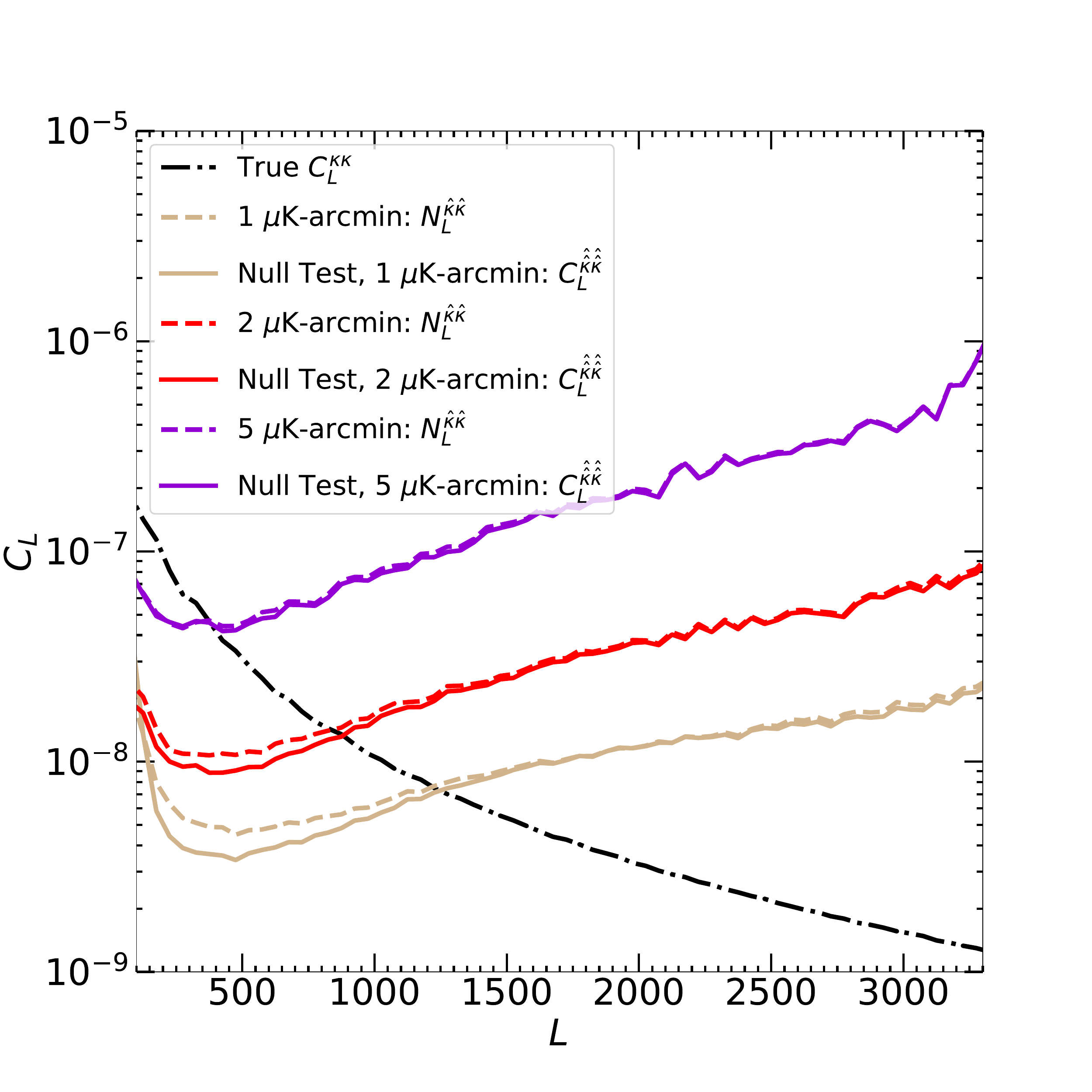}
	\end{center}
	\vspace{-0.4cm}
	\caption{Null test shows  the average power spectra ($C^{\hat{\hat{\kappa}} \hat{\hat{\kappa}}}$) of rescaled lensing maps predicted by the RDLFUnet when we feed unlensed versions of the test set to fully trained network. We compare these rescaled power spectra to the reconstruction noise power spectra ($N_{L}^{\hat{\kappa} \hat{\kappa}}$) of the ResUnet model shown in Figure \ref{recons_noise_power_}.}
	\label{null_test}
\end{figure}

\begin{figure*}
	\begin{center}
		\includegraphics[width=0.8\hsize]{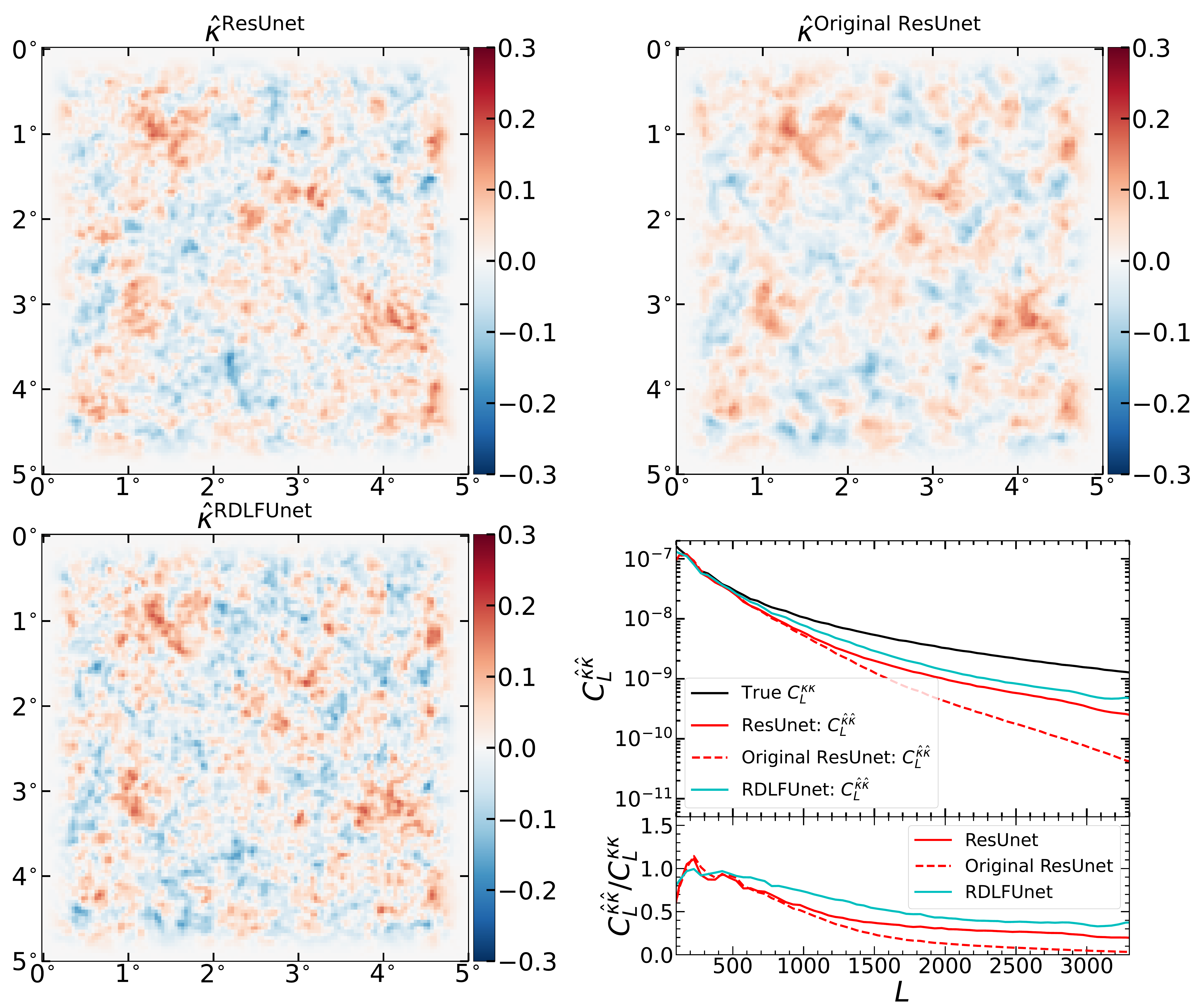}
	\end{center}
	\vspace{-0.4cm}
	\caption{Reconstructed $\kappa$ map using the ResUnet model with FFT loss for the noise level of $1\mu$K-arcmin. As a comparison, reconstructed $\hat{\kappa}$ from the original ResUnet and RDLFUnet also are shown in panels. Here, original ResUnet represents the ResUnet model without FFT loss function. Power spectra of prediction map $\hat{\kappa}$ averaged over all realizations in the test set are also shown in the bottom panel.}
	\label{resunet_map_power}
\end{figure*}

Figure \ref{recons_noise_power_} shows the reconstruction noise power spectrum
for the experiments with  $1 \, \mu$K-arcmin (upper panel) and $5\, \mu$K-arcmin (lower panel) noise levels using the ResUnet model, the standard quadratic estimator, and our network model. The noise power spectra (see Eq. (\ref{qe_NL})) of the quadratic estimator are obtained using  publicly available \texttt{Symlens}\footnote{\url{https://github.com/simonsobs/symlens}} package.

We can see that the ResUnet model predictions have a lower reconstruction noise than the quadratic estimator at all scales for the noise level of $1\ \mu$K-arcmin and closely matches the performance of the quadratic estimator for the noise level of $5\ \mu$K-arcmin. The ResUnet model performance is also obviously better than the quadratic estimator for the noise level of $2\ \mu$K-arcmin (the result is not shown in the figure). These imply that ResUnets outperform the QE method for lower noise levels. These conclusions are consistent with  those made by \cite{Caldeira:2019}.

For experiments with noise levels below $5\ \mu$K-arcmin, the reconstruction noise levels from our network model are lower than those from the ResUNet model and QE at almost all angular scales. These results suggest that our approach outperforms both the QE and the ResUnet model in terms of signal-to-noise ratio on the reconstructed $\kappa$ map.

Finally, we use the Fisher matrix method to compute the signal-to-noise ratio (SNR) \citep{LiuSun:2022}
\begin{align}
 	\label{snr}
 	{\rm SNR} = \sqrt{\sum_{L,L'}C_{L}\mathbb{C}_{L,L'}^{-1}C_{L'}},
\end{align}
where the numerator $C_{L}$ is the theoretical $\kappa$ spectrum, and $\mathbb{C}_{L,L'}$ is the covariance matrix obtained from our test set via
\begin{align}
	\mathbb{C}_{L,L'} = \frac{1}{N-1}\sum_{n=1}^{N=2500}[ (C^{\hat{\kappa} \hat{\kappa}}_{L} - \bar{C}^{\hat{\kappa} \hat{\kappa}}_{L}) \times (C^{\hat{\kappa} \hat{\kappa}}_{L'} - \bar{C}^{\hat{\kappa} \hat{\kappa}}_{L'}) ],
\end{align}
where $C^{\hat{\kappa} \hat{\kappa}}_{L}$ is reconstructed $\kappa$ power spectrum and $ \bar{C}^{\hat{\kappa} \hat{\kappa}}_{L}$ is the averaged reconstructed $\kappa$ power spectrum across the test set. For experiments with the noise level of $1\ \mu$K-arcmin, the SNRs for our network model, ResUNet, and QE estimator are 14.7, 12.4, 9.1, respectively.  For experiments with the noise level of $5\ \mu$K-arcmin, the SNRs for our network model, ResUNet, and QE estimator are 6.1, 5.5, 5.2, respectively. Our method can get SNR higher than ResUNet and QE methods.

\subsection{Null test}
Finally, a null test is performed to ensure that the network learns a sensible mapping of the input lensed ($Q^{\rm obs}$,\,$U^{\rm obs}$) maps to the predicted $\hat{\kappa}$ map. In other words, we need to check that the predicted $\hat{\kappa}$ maps by our network model are due to the presence of the true $\kappa$ signal rather than an artifact of the network (i.e., non-physical signal). In order to check the output results of our network model,  we feed unlensed versions of ($Q$,\,$U$) maps to the network trained on observed ($Q^{\rm obs}$, $U^{\rm obs}$) maps. 

We add noise to unlensed ($Q^{\rm prim}$,\,$U^{\rm prim}$) maps introduced in the section \ref{data_pipline} and feet it to network trained on observed ($Q^{\rm obs}$,\,$U^{\rm obs}$) maps. The cross-correlation spectrum ($C^{\hat{\kappa}\kappa}_L$) of reconstructed $\hat{\kappa}$ derived from the null test and true $\kappa$ is first calculated, and it is discovered that they are uncorrelated at all noise levels. This suggests that when we input the unlensed ($Q$,\,$U$) maps into network trained on observed maps, the output reconstructed $\kappa$ map of the network is merely random noise.

The power spectrum of the output $\hat{\kappa}$ map from the null test is then calculated and compared to the reconstructed noise. To address bias, we first use Eq.(\ref{norm}) to normalize the biased $\hat{\kappa}$ map from the null test. We can calculate the rescaled power spectrum $C^{\hat{\hat{\kappa}} \hat{\hat{\kappa}}}_{L}$ in Eq.(\ref{rescale}) from the output of null test, and compare it to the reconstructed noise shown in Figure \ref{recons_noise_power_}. Figure \ref{null_test} shows the rescaled power spectra $\langle C^{\hat{\hat{\kappa}} \hat{\hat{\kappa}}}_{L} \rangle$  for each noise level and the reconstructed noise. We can see that the rescaled power spectra from the null test are basically consistent with the reconstructed noise at all noise levels, which is also consistent with the result of \citet{Caldeira:2019}. All these null tests illustrate that our network  is capable of successfully learning the reasonable mapping from the input lensed ($Q^{\rm obs}$,\,$U^{\rm obs}$) maps to the predicted $\hat{\kappa}$ map correctly.

We also notice that there is the difference within a few percent between $ C^{\hat{\hat{\kappa}} \hat{\hat{\kappa}}}_{L}$ and $N^{\kappa \kappa}_L$ at large scales, and this difference becomes more obvious for the input maps with lower noise levels. We think that this could be caused by the normalization factor $R$ in Eq.(\ref{Rnorm}), which may be not very accurate and will bias the $\hat{\hat{\kappa}}$ field in computation because the $R$ computed from the training set fluctuates high/low.

\section{Discussions} 
\label{discus}

\subsection{Comparing with ResUnet Methods}
In this work, we compared the performances of the ResUnet and RDLFUnet in reconstructing $\kappa$ map. To be consistent with \citet{Caldeira:2019}, the ResUnet model uses mean squared error in image space as the loss function for training, whereas the RDLFUnet uses Eq.(\ref{loss}). The two models should utilize the same loss function for an accurate comparison.

We use the loss function contained FFT loss, Eq.(\ref{loss}), to train the ResUnet model for the experiment of $1\mu$K-arcmin noise (other experiments have similar results), and results of reconstructing $\kappa$ are presented in Figure \ref{resunet_map_power}. We can see that the reconstructed $\hat{\kappa}$ map by the ResUnet with FFT loss function  has more small-scale structures than the results of the original ResUnet model. This implies that the FFT loss function can help the network capture more small-scale structures at the map level, implying that using the FFT loss function could enhance the model's ability to reconstruct $\kappa$ map.

From the aspect of the reconstructed power spectrum, the reconstructed $C^{\hat{\kappa} \hat{\kappa}}$ power spectrum from the ResUnet with FFT loss obviously outperforms the result from the original ResUnet model, but it is still worse than the result of the RDLFUnet. These demonstrate that the $\kappa$ reconstruction process is aided by our loss function and the RDLFUnet model outperforms the ResUnet model. The RDLFUnet model actually has a more complicated structure than the ResUnet model, which suggests that it may have a stronger capacity for fitting  than the ResUnet model.  As a result, compared with the ResUnet method, both our network structure and loss function  can improve the performance of reconstructing $\kappa$ map.  

\subsection{Variability of noise level}
\begin{figure}
	\begin{center}
		\includegraphics[width=1\hsize]{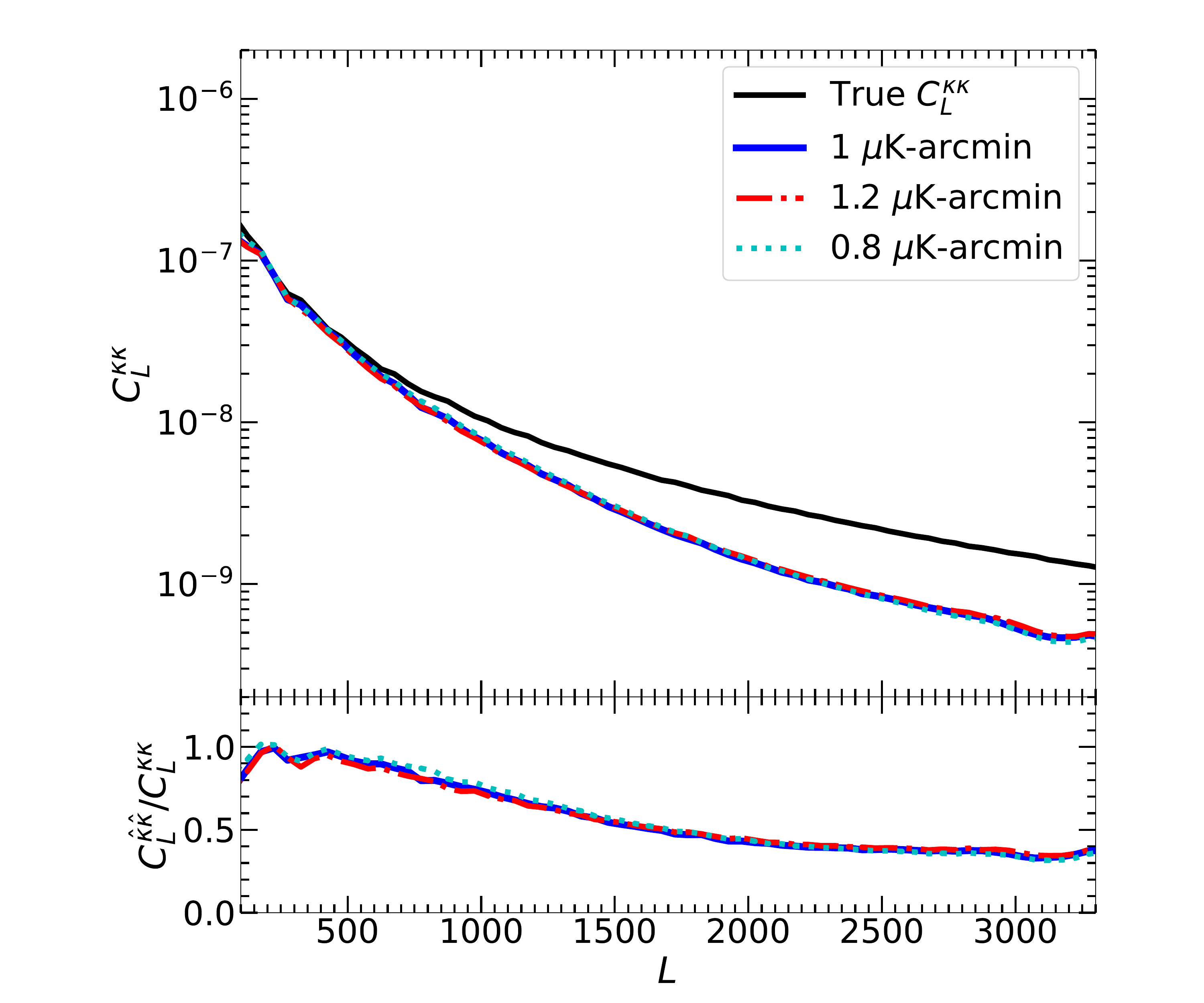}
	\end{center}
	\vspace{-0.4cm}
	\caption{Power spectra of the predicted $\hat{\kappa}$ map from RDLFUnet model.  RDLFUnet model is trained on a training set with noise level of $1\ \mu$K-arcmin. `$1\ \mu$K-arcmin' represents the prediction of RDLFUnet on test sets with noise level of  $1\ \mu$K-arcmin. `$0.8\ \mu$K-arcmin' and `$1.2\ \mu$K-arcmin' are  the prediction of RDLFUnet on test set with noise level of  $0.8\ \mu$K-arcmin and $1.2\ \mu$K-arcmin, respectively. The ratio of the predicted power spectrum to the true power spectrum is also shown.} 
	\label{results_noise1p2}
\end{figure}

In Section \ref{sec_4}, each noise level experiment is trained using a separate network. Here, we alter the noise level for a trained network to see how sensitive our method is to changes in noise level. We train the RDLFUnet model on a training set with noise level of $1\ \mu$K-arcmin  and then feed it three test sets with noise levels of $0.8\ \mu$K-arcmin, $1\ \mu$K-arcmin, and $1.2\ \mu$K-arcmin. We calculate the predicted power spectrum by RDLFUnet model as shown in Figure \ref{results_noise1p2}. The predicted $\hat{\kappa}$ power spectrum for test sets with noise levels of $0.8\ \mu$K-arcmin and $1.2\ \mu$K-arcmin are consistent with the test sets with noise level of $1\ \mu$K-arcmin. This implies that our network is insensitive to small changes in noise level.

\section{CONCLUSIONS}
\label{sec_5}
In this work, we present a machine-learning method for reconstructing the CMB lensing potential. The network was trained using simulated CMB maps with $5 \times 5$ ${\rm deg}^2$ size. We compared RDLFUnet's performance with that of ResUnet model from previous work. We demonstrated that the RDLFUnet can capture more small-scale structure features than ResUnet at the reconstructed $\kappa$ map level based on the reconstructed results of four tests with varying noise levels (0, 1, 2, 5 $\mu$K-arcmin).

At the reconstructed power spectrum level,  we demonstrated that our method outperforms the ResUnet in reconstructing power spectra of lensing convergence and has a higher signal-to-noise ratio than the ResUnet and quadratic estimator methods at almost the entire observation scales.

All results in this paper are based on simulated data. To extend this network usage for actual data, we need to include foreground residual in the input maps. We plan to reconstruct CMB lensing from the foreground cleaned CMB polarization data in future work. In addition,  since actual data are often taken from larger regions of the sky, we also need to use simulations from larger sky patches. As a result, the flat sky approximation will no longer be valid, necessitating the modeling of a spherical sky map. The spherical HEALPix format map needs to be converted into 2D data since the neural network needs 2D data for its inputs. This issue has been addressed in \cite{Wang:2022} and \cite{Yan:2023}. We also investigate  the impact of different foreground removal methods on the results of reconstructing CMB lensing in future work.

We will further investigate the ability of our method to other aspects, such as the reconstruction of foregrounds and CMB delensing, as well as the component separation of future radio surveys, in future works.

\section*{Acknowledgements}
J.-Q. Xia is supported by the National Science Foundation of China under grants No. U1931202 and 12021003; the National Key R\&D Program of China No. 2017YFA0402600 and 2020YFC2201603.



\bibliographystyle{mnras}

\end{document}